\documentclass[dvips]{mn2e}
\usepackage{graphicx}
\title[GRS~1915+105 jet emission model]
{Modelling oscillations in the jet emission from microquasar GRS 1915+105}
\author[R. S. Collins, C. R. Kaiser and S. J. Cox]
{R. S. Collins$^{1,~2}$\thanks{e-mail: rsc@astro.soton.ac.uk}, C. R. Kaiser$^{1}$ and S. J. Cox$^{2}$
\\
$^1$ Department of Physics and Astronomy, University of Southampton, Hampshire, SO17 1BJ, United Kingdom\\
$^2$ Department of Electronics and Computer Science, University of Southampton, Hampshire, SO17 1BJ, United Kingdom}
\begin{document}
\date{Accepted 2002 September 4. Received 2002 July 8}
\pagerange{\pageref{firstpage}--\pageref{lastpage}} \pubyear{2002}

\maketitle
\label{firstpage}

\begin{abstract}

The variability in the infrared to millimetre emission from microquasar GRS 1915+105 is believed to be dominated by the system's relativistic jet. In this paper we develop a time-dependent version of the Blandford \& K\"{o}nigl (1979) jet emission model and apply it to the oscillations in the infrared and millimetre emission from GRS 1915+105 observed by Fender \& Pooley (2000). The resulting model provides a reasonable description of the observed flux oscillations from GRS 1915+105. From a fit of the observed time lag between the flux peaks in the infrared and millimetre emission together with the flux normalisation we were able to determine the model parameters for the GRS 1915+105 jet. We find that to achieve the observed flux levels with the model requires an unphysically large electron density within the jet. We therefore conclude that the Blandford \& K\"{o}nigl (1979) model cannot explain these observations, either because it does not provide the correct description of the emission from microquasar jets, or because the observed emission variations do not originate in the jet.

\end{abstract}


\section{Introduction}\label{sIntro}

The microquasar GRS 1915+105 shows time variability in its emission at all observable wavelengths. Its high energy emission (X-rays) is believed to be created in the inner radii of the system's accretion disc, and its low energy (radio) emission is primarily from synchrotron emitting ejecta located at large radii ($r > 10^{11}$ m) within the system's jet. In this paper we investigate the cause of the variability in the infrared, and millimetre emission from GRS 1915+105 which is believed to originate in the inner regions of the relativistic jet. 

Observations ranging from radio to infrared frequencies have revealed very large amplitude, quasi-periodic oscillations in the emission from GRS 1915+105 (Pooley \& Fender 1997; Mirabel et al. 1998; Fender \& Pooley 1998; Fender et al. 2002). In this paper we will concentrate on the oscillations detected at millimetre and infrared frequencies by Fender \& Pooley (2000). These oscillations are characterised by the short time lags between flux peaks and a flux ratio, between these two frequencies, that is close to unity. This suggests a flat spectrum over a frequency range that encompasses three orders of magnitude. Although flat spectra had previously been observed from GRS 1915+105, these measurements were the first to suggest they extend to the infrared.

Flat spectra from compact radio sources such as microquasars have traditionally been explained in terms of partially self-absorbed synchrotron emission from the inner-most regions of a narrow, conical, relativistic jet (originally by Blandford \& K\"{o}nigl 1979, hereafter referred to as BK79, but also by Hjellming \& Johnston 1988, Falcke \& Biermann 1995, Falcke 1996, and Falcke \& Biermann 1999). This interpretation is supported by radio observations of blobs moving at apparent superluminal velocities outwards from a central elongated radio core (Mirabel \& Rodr\'{\i}guez 1994; Rodr\'{\i}guez \& Mirabel 1999; Fender et al. 1999; Dhawan, Mirabel \& Rodr\'{\i}guez 2000). Detection of polarised emission strongly suggests the synchrotron nature of the emission (e.g. Fender et al. 2002). The temporal and spectral behaviour of the emission coming from the blobs is consistent with expectations from internal shocks in a relativistic jet flow (Kaiser, Sunyaev \& Spruit 2000). These synchrotron-emitting blobs are therefore believed to be ejecta from a central compact, relativistic jet.

However, the BK79 model only represents a steady-state solution for emission from relativistic jets. Using the basic principles of the BK79 model we develop a time-dependent variation of this model in an attempt to explain the observed infrared and millimetre flux oscillations from GRS 1915+105. Whereas previous time-dependent jet emission models have only been concerned with the emission from the spherical ejecta blobs (e.g. van der Laan 1966, and Hjellming \& Johnston 1988), the model presented in this paper concentrates solely upon the variability in the emission from the inner-most radii of the conical jet. In section \ref{sBKModel} we demonstrate both how the steady-state BK79 model produces a flat synchrotron spectrum and how this is dependent upon the physical parameters of the jet. The model is then adapted to include time variability in section \ref{sBKTEModel}, which allows a direct comparison between the observations and the model in section \ref{sObs}. Finally we discuss the successes and failures of the model in explaining the observations in section \ref{sDis}.

%------------------------------------------------------------------------------
%****************************************************************************** %------------------------------------------------------------------------------

\section{The Blandford-K\"{o}nigl steady-state jet model}\label{sBKModel}

Observations of flat spectrum synchrotron radiation are explained by the BK79 model as the emission from a supersonic, relativistic jet at radii relatively close to its acceleration region. The jet has a conical geometry, with a constant opening angle, as a consequence of the free adiabatic expansion of the jet plasma which is modelled as flowing at a constant velocity, $v_{\rm j}$. Irregularities in the supersonic flow gives rise to shocks that accelerate some of the electrons in the plasma to relativistic velocities with a power-law energy distribution of
\begin{equation}
\label{eN}
n(\gamma) = k\gamma^{-p},
\end{equation}
up to a maximum value of $\gamma_{\mathrm{max}}$, where $\gamma$ is the electron energy in terms of the Lorentz factor, $k$ is a normalisation value for the number density, and $p$ the power-law index (typically $\sim 2$ from acceleration by shocks, e.g. Bell, 1978). 

Flux conservation requires that the magnetic field strength, $B$, in this conical jet varies as $r^{-1}$, where $r$ is the radial distance from the origin. The radial dependence of the electron density normalisation value is solely determined by volume expansion in two dimensions and so $k$ varies as $r^{-2}$. This represents the most important assumption of the BK79 model for the creation of flat emission spectra. It implicitly proposes the existence of ongoing particle acceleration throughout the expanding jet to compensate for the energy losses of the relativistic electrons due to adiabatic decompression. Furthermore, radiative losses due to synchrotron emission and inverse Compton scattering of the Cosmic Microwave Background (CMB) are also suppressed, allowing the maximum electron energy cut-off at $\gamma_{\mathrm{max}}$ to remain constant. Thus the power-law distribution of $n(\gamma)$ holds for all times (hence the {\em steady-state} jet model) throughout the jet. Although, as stated by BK79, this assumption is a requirement for the equipartition of energy between the magnetic field and the electrons, the cause of the ongoing particle acceleration is unclear as a viable physical mechanism has not been identified. Note that this re-acceleration process differs from that which is usually attributed to shock-wave propagation in jets, in that it must be a continuous, ubiquitous process that reacts to energy losses in such a way as to exactly counter the losses.

The jet's emission is entirely due to self-absorbed synchrotron radiation, with emission, $\epsilon_{\nu}$ (J s$^{-1}$ m$^{-3}$ Hz$^{-1}$ sr$^{-1}$), and absorption, $\chi_{\nu}$ (m$^{-1}$), co-efficients that for a power-law distribution of electrons are given by
\begin{eqnarray}
\label{eEmCo}
\epsilon_{\nu} & = & \frac{\sqrt{3\pi}\,e^3}{64 \pi^3 \epsilon_0 m_{\mathrm{e}} c\,(p+1)} 
  \biggl(\frac{3e}{2\pi\,m_{\mathrm{e}}}\biggr)^{\!(p-1)/2} 
\nonumber \\   &   &  \times \; \frac{\Gamma\left[(3p+19)/12\right]
  \Gamma\left[(3p-1)/12\right]\Gamma\left[(p+5)/4\right]}
  {\Gamma\left[(p+7)/4\right]} \nonumber \\
               &   & \times \; k \; B^{(p+1)/2} \; \nu^{-(p-1)/2}
\end{eqnarray}
and
\begin{eqnarray}
\label{eAbsCo}
\chi_{\nu} & = & \frac{\sqrt{3\pi}\,e^3}{64 \pi^2 \epsilon_0 m_{\mathrm{e}}^{\,2} c} 
  \left(\frac{3e}{2\pi\,m_{\mathrm{e}}}\right)^{\!p/2} 
  \; k \; B^{(p+2)/2} \; \nu^{-(p+4)/2} \nonumber \\                     
           &   & \times \; \frac{\Gamma\left[(3p+22)/12\right]                     \Gamma\left[(3p+2)/12\right]\Gamma\left[(p+6)/4\right]}
  {\Gamma\left[(p+8)/4\right]} 
\end{eqnarray}
respectively (e.g. Longair 1994). When applied to the jet geometry, where $k = k(r)$ and $B = B(r)$, the source function varies with radius as
\begin{equation}
\label{eSourceP}
S_{\nu}(r) = \frac{\epsilon_{\nu}(r)}{\chi_{\nu}(r)}
 \propto r^{1/2} \; \nu^{5/2}.
\end{equation}
For a jet observed at an inclination of $90^\circ$, the path length along the line-of-sight is assumed to be equal to the jet's half-width, $w$, which for a conical jet varies as $r$. Therefore the optical depth varies with radius as
\begin{equation}
\label{eDepthP}
\tau_{\nu}(r) = \chi_{\nu}(r) w(r)
 \propto r^{-(p+4)/2} \; \nu^{-(p+4)/2}.
\end{equation}

To obtain the emission spectrum the equation for radiative transfer through a homogenous medium is solved, which gives a specific intensity of
\begin{equation}
\label{eI}
I_{\nu} = S_{\nu} \left(1 - \mathrm{e}^{-\tau_{\nu}}\right).
\end{equation}
The observed flux density, $F_v$ (W m$^{-2}$ Hz$^{-1}$), is then simply calculated as the product of the specific intensity with the projected surface area of the jet, 
\begin{eqnarray}
\label{edF}
\mathrm{d}F_{\nu} & = & I_{\nu} \; \mathrm{d}\Omega \\
\mathrm{d}\Omega & = & \frac{2 \pi w \mathrm{d}r}{{D_\mathrm{j}}^2}, \nonumber
\end{eqnarray}
where $D_\mathrm{j}$ is the distance to the jet. Hence
\begin{equation}
\label{edF2}
\mathrm{d}F_{\nu} \propto r^{3/2} \nu^{5/2} \left[1 - \mathrm{e}^{-r^{-(p+4)/2} \; \nu^{-(p+4)/2}}\right] \,\mathrm{d}r.
\end{equation}

%------------------------------------------------------------------------------

\subsection{Analytical solution}\label{ssBKModel_Analytical}

A flat emission spectrum results from integrating the emitted flux from the base of the jet out to the maximum observable radius, $r_{\mathrm{max\;ob}}$, where the magnetic field strength is too weak to produce synchrotron radiation at a given frequency. This radius is dependent upon the critical frequency, $\nu_{\mathrm{c}}(r) \propto \gamma_{\mathrm{max}}^2 B(r)$, of those electrons with the maximum Lorentz factor, $\gamma_{\mathrm{max}}$, which remains constant as a result of the postulated re-acceleration process. If the critical frequency of these most energetic electrons falls below the observing frequency, $\nu$, then the maximum radius from which we receive radiation of frequency $\nu$ is reached. Hence $r_{\mathrm{max\;ob}} \propto \nu^{-1}$ for a conical geometry. The flux integration may be approximated by separating the optically thin and optically thick regions of the jet,

\begin{eqnarray}
\label{eFP}
F_{\nu} & \propto & \int_{r_{\mathrm{th}}(\nu)}^{r_{\mathrm{max\;ob}}} r^{3/2} \; \nu^{5/2} \times r^{-(p+4)/2} \; \nu^{-(p+4)/2} \;\mathrm{d}r \nonumber \\
        &         & + \int_{0}^{r_{\mathrm{th}}(\nu)} r^{3/2}\;\nu^{5/2}\;\mathrm{d}r \\
        & \propto &  \nu^0, \nonumber
\end{eqnarray}
where the optically thin radius, $r_{\mathrm{th}}(\nu) \propto \nu^{-1}$, is defined as the radius where the optical depth of the jet material for a given frequency, $\nu$, is equal to unity. This recovers the flat spectrum result of BK79 for the steady-state scenario.

%------------------------------------------------------------------------------

\subsection{Numerical study}\label{ssBKModel_Numerical}

When modelling time variability, in \S \ref{sBKTEModel}, the emission region is constrained to some fixed length, terminating at a radius determined by the properties of the jet flow rather than the observing frequency. In this case a flat spectrum can also be created, without the requirement that the maximum observable radius, $r_{\rm max\;ob}$, lies within the emission region. However, the flat region of the spectrum will only extend over a finite frequency range, with an optically thick, $F_{\nu} \propto \nu^{5/2}$, slope at the low frequency cut-off, and an optically thin, $F_{\nu} \propto \nu^{-(p-1)/2}$, slope at the high frequency cut-off. 

\begin{figure}
\centering
\includegraphics[width=8cm]{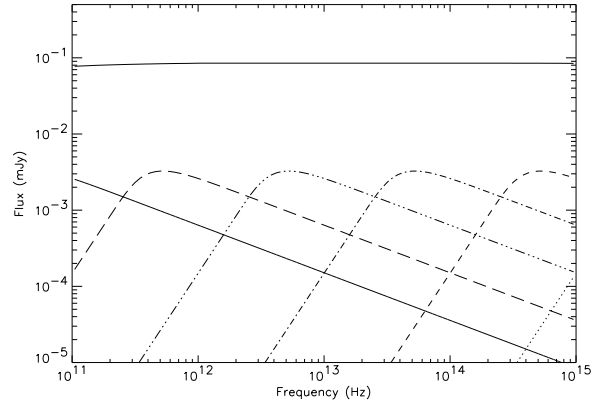}
\caption{The formation of a flat spectrum from the summation of self-absorbed synchrotron spectra. Here a sample of spectra emitted from electron populations located at increasing radii along the jet reveal the shift of the individual spectra towards lower frequencies. The higher the frequency of the spectral peak, the closer to the jet origin.}
\label{fFlatSpectrum}
\end{figure}

The flat region of the spectrum encompasses every frequency for which the jet material, over the length of the emission region, is optically thick at the smallest radius and has become optically thin at the largest radius. Thus the flat region is formed by the summation of self-absorbed synchrotron spectra created by electron populations at increasing radii along the jet. As demonstrated in figure \ref{fFlatSpectrum}, each individual spectrum is shifted towards lower frequencies as the electron density and magnetic field strength decrease at greater radii. If energy losses due to adiabatic decompression are included in the model, then the spectra with lower frequency peaks (from greater radii) will have decreasing flux levels, hence an inverted spectrum is observed rather than a flat spectrum.

Therefore by suitably adjusting the parameters that control the optical depth of the jet material as a function of radius, the frequency range of the flat region of the spectrum, extending from $\nu_{\rm min}$ to $\nu_{\rm max}$, may be shifted to agree with observations. However, the extent of the frequency range, $\nu_{\rm max} / \nu_{\rm min}$, is solely determined by the length of the emission region. In this section we demonstrate these principles by numerically integrating equation \ref{edF}, which calculates the flux over the length of the jet.

The emission and absorption co-efficients as function of radius are calculated from equations \ref{eEmCo} and \ref{eAbsCo}, with the magnetic field strength, $B(r) = B_0 (r/r_0)^{-1}$, and the normalisation value of the electron density distribution, $k(r) = k_0 (r/r_0)^{-2}$. Here $B_0 = B(r\!=\!r_0)$, and $k_0 = k(r\!=\!r_0)$. As the power-law index of the electron distribution, $p$, is commonly stated to lie between 2.2 and 2.3 for relativistic shock acceleration (e.g. Achterberg et al. 2001), we fix $p$ to a value 2.25. The precise value of $p$ is unimportant in practice as when varied over the range $2 < p < 3$ it only has a negligible effect upon the model results. Hence the source and optical depth functions become
\begin{eqnarray}
\label{eSource}
S_{\nu}(r) & = & 1.3 \times 10^{-14} \; B_0^{-1/2}
  \left(\frac{r}{r_0}\right)^{1/2} \left(\frac{\nu}{\rm GHz}\right)^{5/2}
  \nonumber \\ 
           &   & \mbox{ J s$^{-1}$ m$^{-2}$ Hz$^{-1}$ sr$^{-1}$},
\end{eqnarray}
and
\begin{equation}
\label{eDepth}
\tau_{\nu}(r) = 1.5 \times 10^{-12} k_0 B_0^{17/8} w_0 
\left(\frac{r}{r_0}\right)^{-25/8} \left(\frac{\nu}{\rm GHz}\right)^{-25/8},
\end{equation}
where, for a conical jet, we have $w(r) = w_0 (r/r_0)$. 

The emission spectrum can then be calculated from equation \ref{edF}, by performing a Romberg numerical integration (e.g. Press et al. 1992) of
\begin{eqnarray}
\label{eFN}
F_{\nu} & = & \frac{8.8 \times 10^{-18}}{({D_{\rm j}}/\mathrm{pc})^2} 
  \; B_0^{-1/2} \; w_0 \left(\frac{\nu}{\rm GHz}\right)^{5/2} \nonumber \\
        &   & \times \int_{r_{\rm min}}^{r_{\rm max}}
   \left(\frac{r}{r_0}\right)^{3/2} 
   \left[1 - \mathrm{e}^{-\tau_{\nu}(r)}\right] \,\mathrm{d}r \mbox{ mJy},
\end{eqnarray}
where all units are in SI unless stated otherwise. For a fixed emission region length, and jet opening angle (which determines $w_0$), the model has two parameters; $k_0$, and $B_0$. To simplify the analysis of the model, we introduce a single parameter, $a_0 = k_0 \, B_0^{17/8}$, that uniquely determines the optical depth function. If the flux is normalised, then $a_0$ becomes the only free model parameter, as the other model parameter, $B_0$, is determined by the flux normalisation.

\begin{figure}
\centering
\includegraphics[width=8cm]{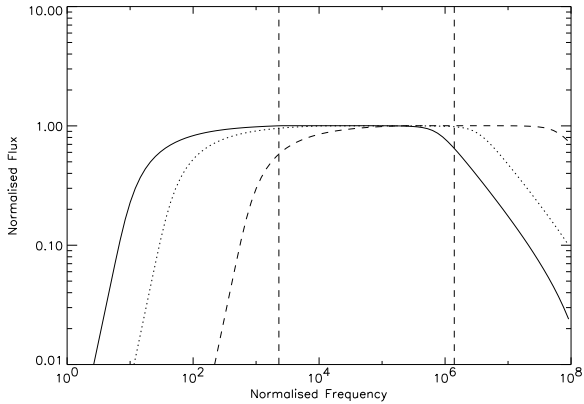}
\caption{Normalised spectra from emission region of length $r_{\rm max} = 10^5 \; r_{\rm min}$ for values $a_0 = 10^2$ (solid line), $a_0 = 10^4$ (dotted line), $a_0 = 10^8$ (dashed line), in SI units. The vertical dashed lines denote the frequencies $\nu_1$, and $\nu_2$, where $\nu_2 > \nu_1$, which are used in section \S \ref{sBKTEModel}}
\label{fss_spectrum}
\end{figure}

The spectra, normalised to the maximum flux, from an emission region of size, $r_{\rm max} / r_{\rm min} = 10^5$, for three different values of $a_0$ are shown in figure \ref{fss_spectrum}. This figure demonstrates that an increase in the value of $a_0$, which increases the optical depth of the jet material, will shift the spectrum towards higher frequencies. An increase in the value of $a_0$ will also increase the unnormalised flux for all frequencies.

The low frequency cut-off to the flat region of the spectrum occurs at the frequency at which the jet material at $r_{\rm max}$ becomes optically thin, so that the entire jet is optically thick to all lower frequencies. Conversely, the high frequency cut-off occurs at the frequency at which the jet material at $r_{\rm min}$ becomes optically thin, so that the entire jet is optically thin to all higher frequencies. Hence, the extent of the frequency range covered by the flat part of the spectrum for a given emission region can be determined by calculating the radii where $\tau_{\nu} = 1$ as a function of frequency. This is described by the function, $r_{\rm th}(\nu) \propto \nu^{-1}$, and hence an emission region of size, $r_{\rm max} / r_{\rm min} = 10^5$, should create a flat spectrum over the frequency range, $\nu_{\rm max} / \nu_{\rm min} = 10^5$. However, as figure \ref{fss_spectrum} demonstrates, the flat region does not remain truly flat over the full span of this frequency range. This is because there is a smooth transition from the flat region of the spectrum to the optically thick and thin power-law tails.

The observations of the GRS 1915+105 jet which are modelled in section \S \ref{sObs}, reveal a flux ratio between two frequencies that is close to, but not equal to, unity. If this slight deviation from a truly flat spectrum is interpreted as the effect of the optically thick cut-off slope, then for a fixed emission region size, the model parameter, $a_0$, may be determined by fitting the model to the observed flux ratio. The flux normalisation may then be used to determine $B_0$, and hence $k_0$ may also be extracted.

%------------------------------------------------------------------------------
%****************************************************************************** %------------------------------------------------------------------------------

\section{Time variability in flat synchrotron emission spectra}
\label{sBKTEModel}

In this section we investigate the possibility of explaining the time variability observed in jet sources with a time-dependent version of the BK79 model. The GRS 1915+105 observations (see \S \ref{sObs}) show that during an oscillation the ratio between the infrared and millimetre flux remains approximately constant, indicating that the flat spectrum apparently persists throughout the full cycle of flux increase and subsequent decrease. Hence the time variability simply represents an oscillation in the flux level of the flat spectrum, without any significant changes in its form over this frequency range. To achieve an increase in the flux normalisation of the flat spectrum, in terms of the steady-state BK79 model, would require a near-simultaneous increase in either the electron density or the magnetic field strength throughout the entire length of the emission region. 

The length of the emission region is determined by the distance the ejected jet material has travelled in the time period defined by the flux oscillation. Emission from material that was ejected at earlier times, and hence is at greater radii, does not need to be modelled as it merely represents a constant flat background component which has no effect upon the time variability. Furthermore, the flux contribution of this background component to the flux peaks is insignificant, as this material will be located at radii where it is very optically thin to radiation of the observed frequency (see figure \ref{fFlatSpectrum}).

From the numerical study of the previous section, the simplest mechanism to mathematically increase the flux normalisation of the flat spectrum would be a decrease in the magnetic field strength. However that requires $a_0$ to remain constant to ensure the flat spectrum still encompasses the same spectral range and hence the electron density would have to increase by an appropriate amount. It is not clear which physical process could cause such a fine-tuned change, and therefore the flux increase is best thought of as an increase in $a_0$, due to an electron density increase and/or a magnetic field strength increase.

In the absence of any clear physical expectations for time variability of the jet properties, and without loss of generality, we interpret the flux increase at the beginning of an oscillation as an increase in the electron density of the material injected into the jet. To simplify the model we begin and end with an injected electron density of zero, although in reality an increase from some quiescent electron density injection value is more likely. The resulting spectral behaviour of the model is the same, but in the latter case our results would just be superimposed upon a pre-existing steady flat spectrum. This underlying flat spectrum would be more extended then that which can vary in the observed time-scale, but that is not of any importance as it has not been observed.

By allowing $k_0 = k_0(t)$, we introduce a time dependence to the observed flux function. However, if the injected electron density at the base of the jet varies with time, then it will also vary with radius, as there is a lag in the time for the injected electrons to reach greater radii along the jet. Therefore we introduce a transformed variable for $t$
\begin{equation}
\label{eTransTime}
t^{\prime}(t, r) = t - \frac{r - r_{\rm min}}{v_{\rm j}}.
\end{equation}
Hence equation \ref{eFN} becomes
\begin{eqnarray}
\label{eFT}
F_{\nu} & = & \frac{8.8 \times 10^{-18}}{({D_{\rm j}}/\mathrm{pc})^2} 
  \; B_0^{-1/2} \; w_0 \left(\frac{\nu}{\rm GHz}\right)^{5/2} \nonumber \\
        &   & \times \int_{r_{\rm min}}^{r_{\rm max}(t)}
   \left(\frac{r}{r_0}\right)^{3/2} 
   \left[1 - \mathrm{e}^{-\tau_{\nu}(t, r)}\right] \,\mathrm{d}r \mbox{ mJy},
\end{eqnarray}
with,
\begin{eqnarray}
\label{eDepthT}   
\tau_{\nu}(t, r) & = & 1.5 \times 10^{-12} \; k_0^{\prime}(t^{\prime}) 
  \; B_0^{17/8} \; w_0 \nonumber \\ 
                 &   & \times \left(\frac{r}{r_0}\right)^{-25/8}
  \left(\frac{\nu}{\rm GHz}\right)^{-25/8}, 
\end{eqnarray}
where $r_{\rm max}(t) = r_{\rm min} + v_{\rm j} t$. Here we have assumed a constant bulk velocity of the jet material, $v_{\rm j}$.

We implemented three variations of the electron density injection function, $k_0^{\prime}(t)$, with $t$ expressed in units of $t_{\rm max}$ -- the time of the maximum injected electron density:

\begin{figure*}
\centering
\includegraphics[width=7.6cm]{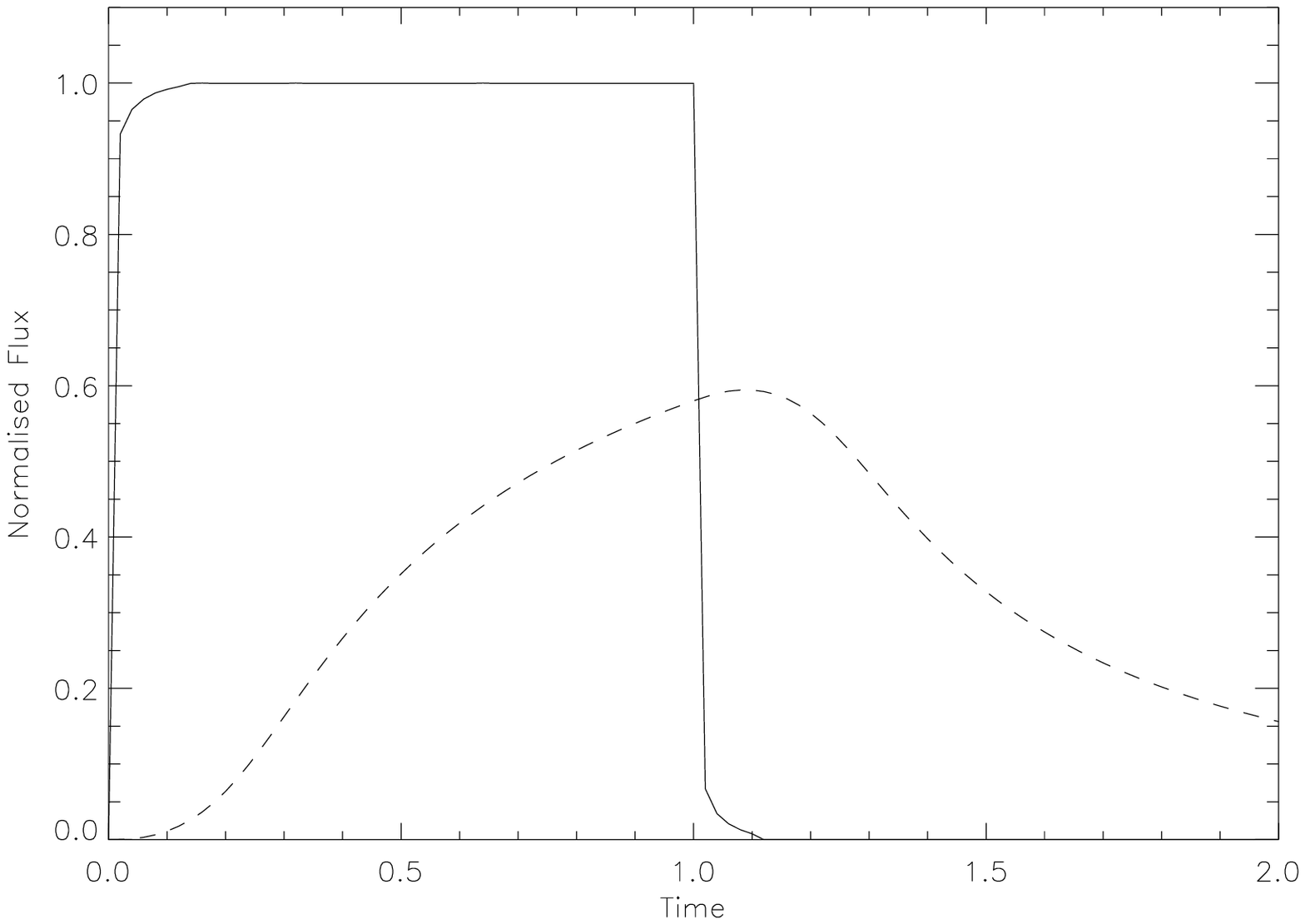}
\includegraphics[width=7.6cm]{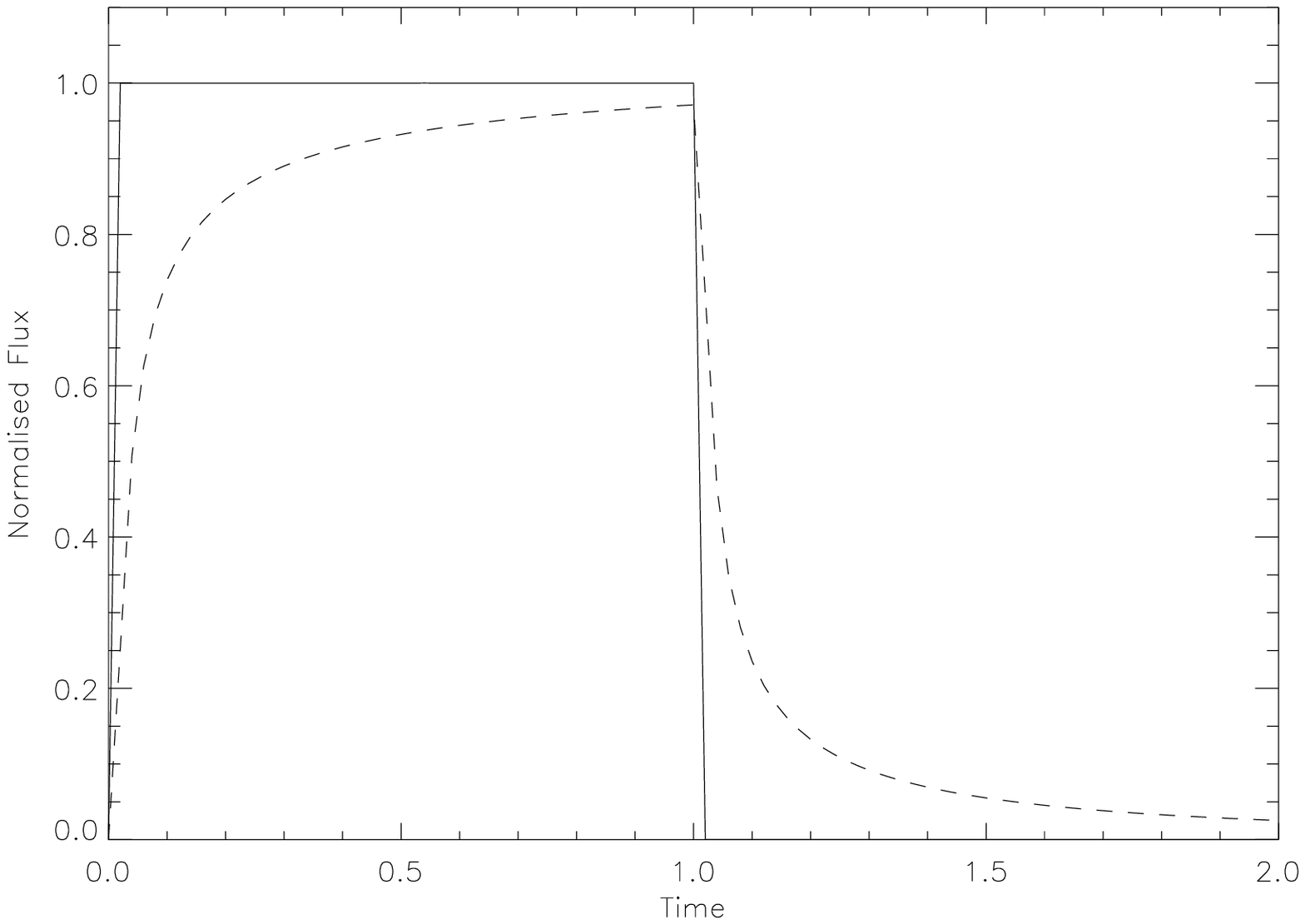}
\caption{Model light curves for $\nu_2$ (solid line) and $\nu_1$ (dashed line), where $\nu_2 > \nu_1$, over a period of $2 \, t_{\rm max}$, for a) $a_0 = 1.0 \times 10^8$ and b) $a_0 = 1.0 \times 10^4$, with a constant value of injected electron density.}
\label{fLCconstant}
\includegraphics[width=7.6cm]{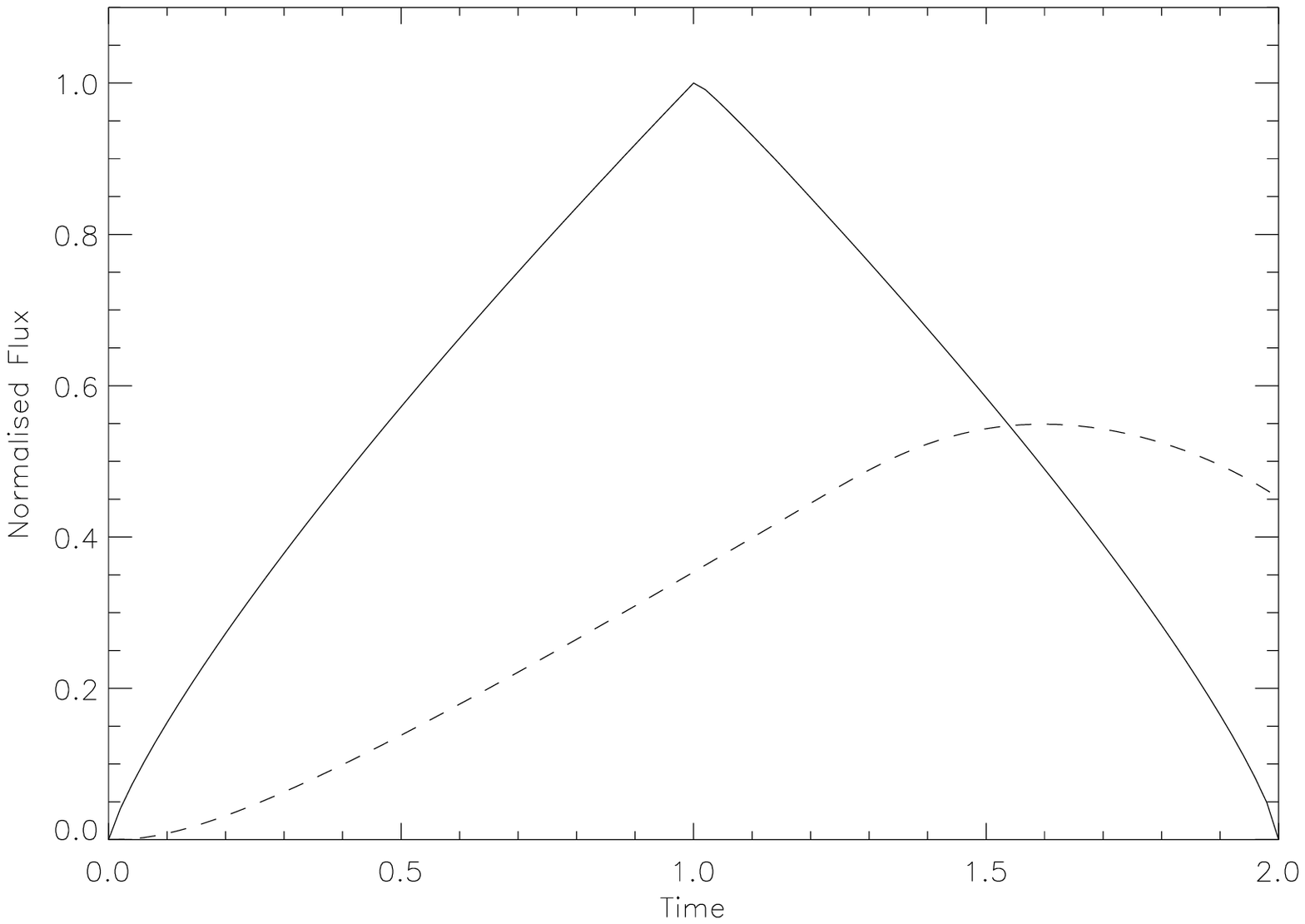}
\includegraphics[width=7.6cm]{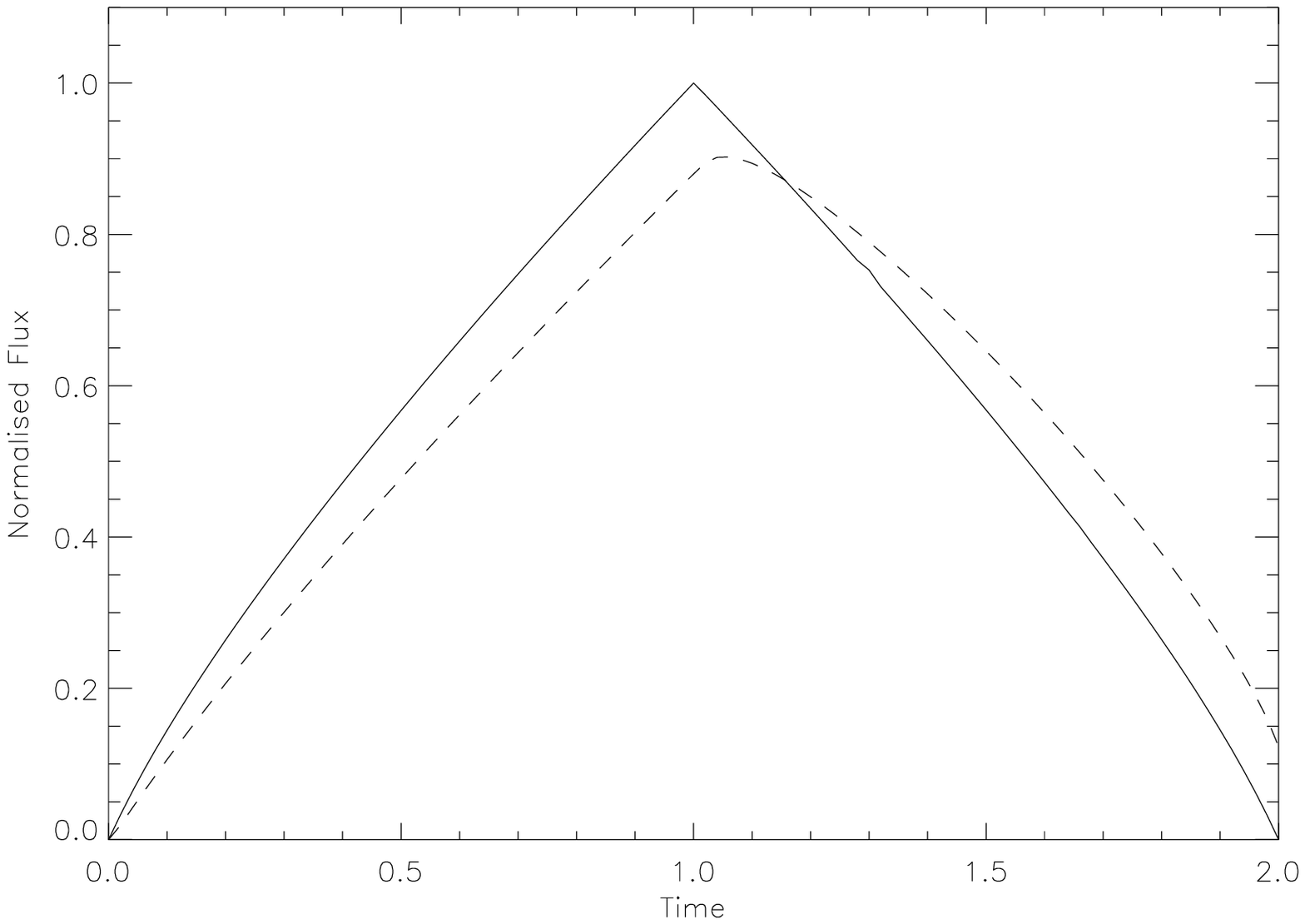}
\caption{Model light curves for $\nu_2$ (solid line) and $\nu_1$ (dashed line), where $\nu_2 > \nu_1$, over a period of $2 \, t_{\rm max}$, for a) $a_0 = 1.0 \times 10^8$ and b) $a_0 = 1.0 \times 10^4$, with a linear electron density injection profile.}
\label{fLClinear}
\includegraphics[width=7.6cm]{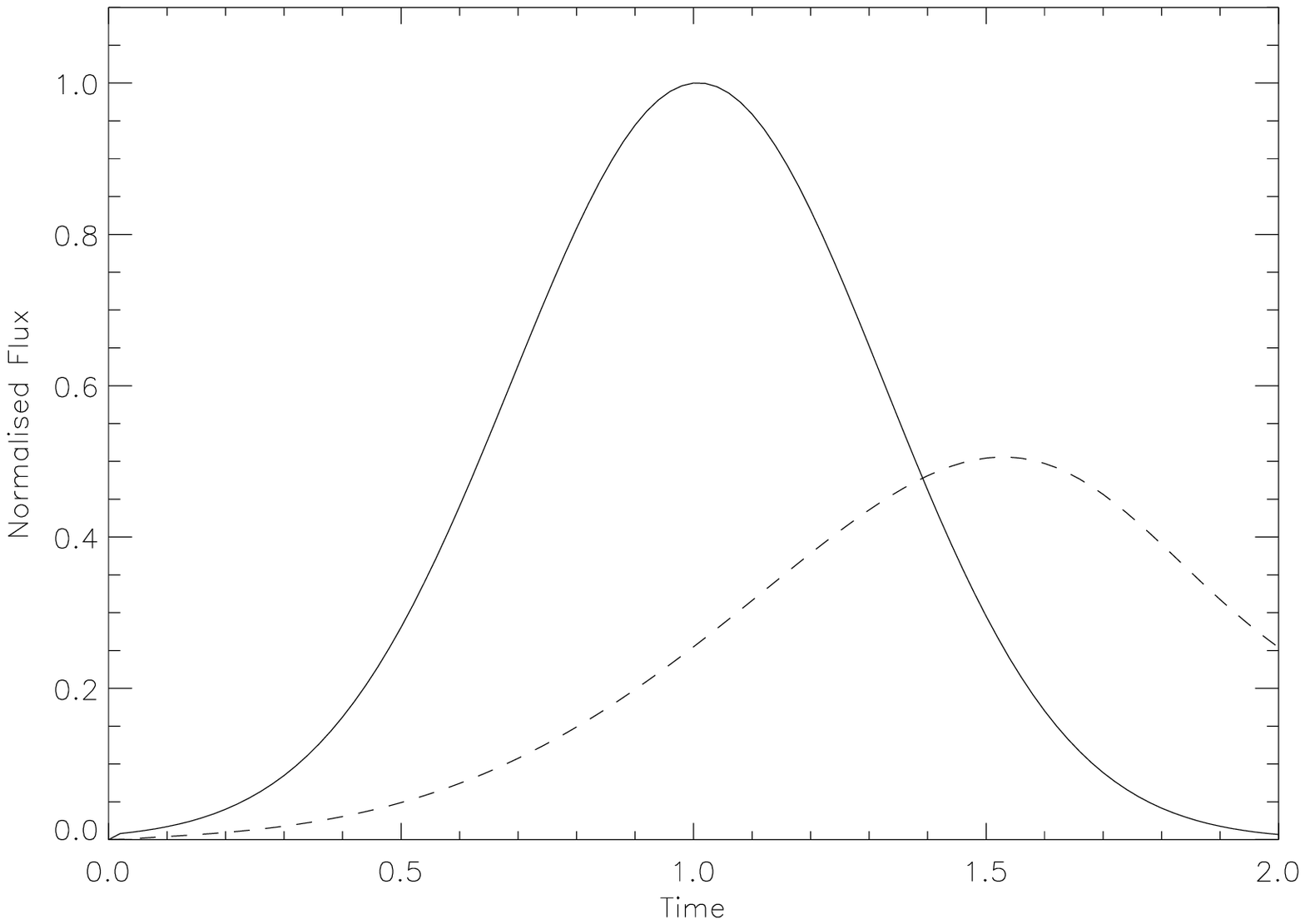}
\includegraphics[width=7.6cm]{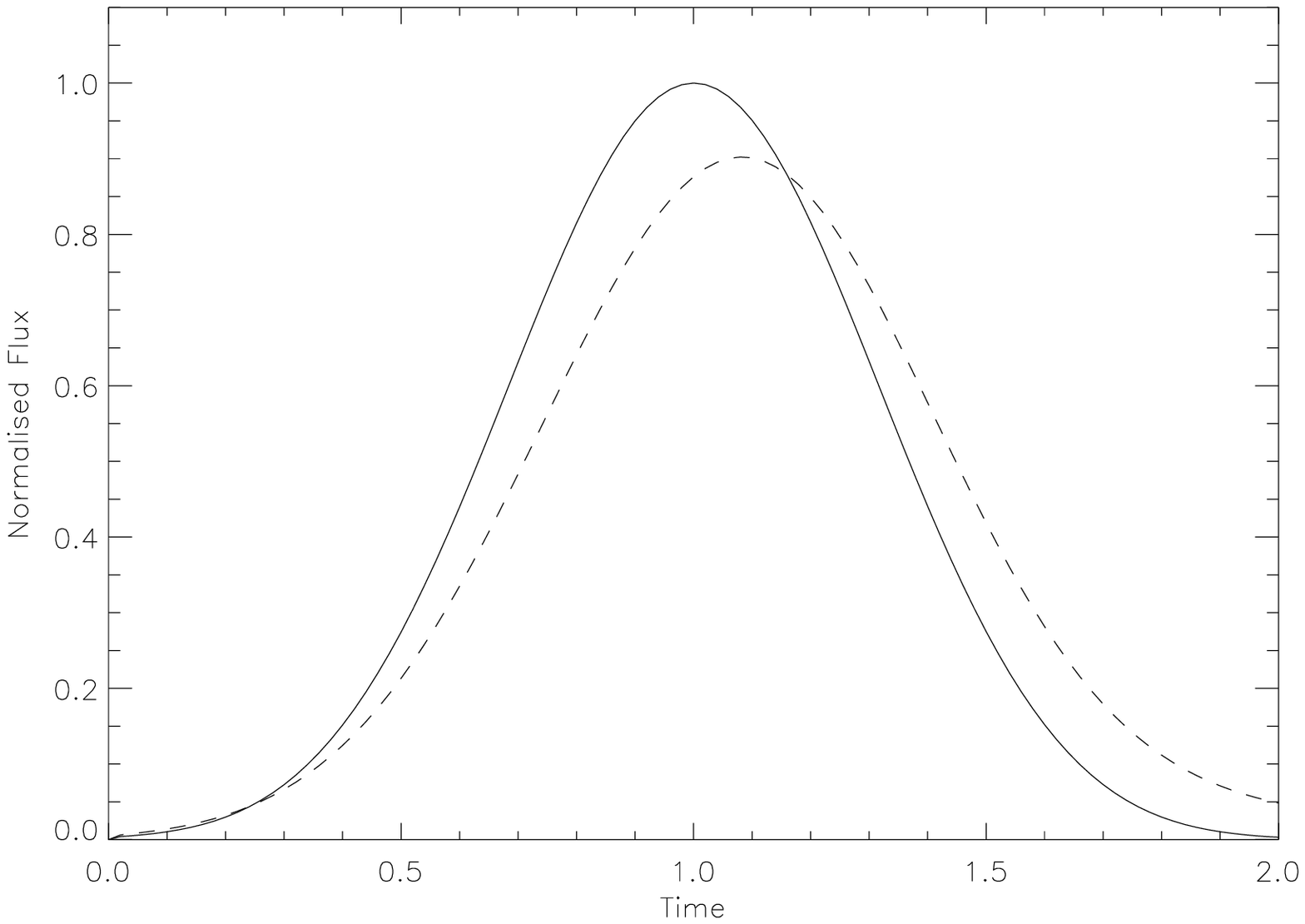}
\caption{Model light curves for $\nu_2$ (solid line) and $\nu_1$ (dashed line), where $\nu_2 > \nu_1$, over a period of $2 \, t_{\rm max}$, for a) $a_0 = 1.0 \times 10^8$ and b) $a_0 = 1.0 \times 10^4$, with a Gaussian electron density injection profile.}
\label{fLCgaussian}
\end{figure*}

1) A simple spontaneous increase in the injected electron density, followed by a spontaneous decrease
\begin{equation}
\label{ek0_constant}
k_0^{\prime}(t) = \left\{ \begin{array} {r@{\quad;\quad}l}
         k_0 & 0 \leq t \leq 1 \\
         0.0 & t > 1          
         \end{array} \right. .  
\end{equation}
This model will effectively produce the steady-state model spectra of figure \ref{fss_spectrum} at, and only at, $t = 1$.

2) A linear increase in the injected electron density, followed by a linear decrease
\begin{equation}
\label{ek0_linear}
k_0^{\prime}(t) = k_0 \left(1 - |t - 1|\right).  
\end{equation}

3) A Gaussian electron density injection function
\begin{equation}
\label{ek0_gaussian}
k_0^{\prime}(t) = k_0 \; \mathrm{e}^{-25(t - 1)^2}.  
\end{equation}

Simulations were performed from $t = 0$ to $t = 2 \,t_{\rm max}$, where 
the values of $t_{\rm max}$ and $v_{\rm j}$ are chosen such that the emission region is the same length as the one used to create the steady-state spectrum in figure \ref{fss_spectrum}. The resulting light curves from two frequencies, $\nu_{\rm 2}$ and $\nu_{\rm 1}$, where $\nu_{\rm 2} > \nu_{\rm 1}$ (as defined in figure \ref{fss_spectrum}), for each of the three injection functions are displayed in figures \ref{fLCconstant}, \ref{fLClinear}, and \ref{fLCgaussian}, respectively. Each injection function light curve pair is shown for two different values of the model parameter $a_0$, which is defined in section \ref{ssBKModel_Numerical}. 

As the increase in electron density propagates along the jet, a flat spectrum, which initially forms at the high frequency end, extends towards lower frequencies. The light curves peak at the time when the flat spectrum has extended to include the frequency of observation. Hence, the light curve of the higher frequency emission always peaks before that of the lower frequency emission because it originates from smaller radii within the jet. The profile of the light curve in figure \ref{fLCconstant} clearly demonstrates how a flat spectrum is created as a function of time. The flat region of the spectrum is formed very rapidly at the high frequency end, but its progression towards lower frequencies slows down over time. When the electron injection is switched off, the high frequency end of the flat spectrum very rapidly disappears but the lower frequency region decays more gradually.

For higher values of $a_0$, the flat spectral region is shifted towards higher frequencies, so a given frequency peaks at a later time. Furthermore, for the lower frequencies, such as $\nu_1$ in figure \ref{fLCconstant}, and for lower values of $a_0$, the flat spectral region never reaches the frequency of observation, and so a slow flux increase, followed by a slow decay is observed. Conversely, as the observation frequency approaches the optically thin cut-off frequency, the observed light curve profile will tend towards the injection function profile. This may also be described by the optically thin radius for the frequency of observation, $\nu$, approaching the base of the emission region, $r_{\rm th}(\nu) \rightarrow r_{\rm min}$.

The linear and Gaussian injection functions both display light curve profiles, shown in figures \ref{fLClinear} and \ref{fLCgaussian}, that tend towards the injection profile as $r_{\rm th}(\nu) \rightarrow r_{\rm min}$. The time lag observed between the flux peaks of the two frequencies can be reduced by decreasing $a_0$, such that both frequencies lie closer to the, rapidly formed, high frequency end of the flat spectrum. This also has the effect of decreasing the peak flux ratio between the two frequencies, until both frequencies lie in the flat region where it becomes unity (assuming the emission region is sufficiently large). For sufficiently small values of $a_0$ the peak flux ratio will begin to decrease again, as the higher frequency emission becomes optically thin. 

For any fixed value of $a_0$ the predicted time lags, peak flux ratios, and flux normalisations are the same for both the linear and Gaussian injection functions, and this would be true for all injection functions with some gradual increase and decrease. Therefore for a given set of frequencies, and a given period of flux increase, $t_{\rm max}$, a value of $a_0$ may be determined from the observed time lags and peak flux ratios, independent of the exact functional form of $k_0^{\prime}(t)$. The flux normalisation then determines $B_0$, and hence $k_0$ may be extracted. This technique is employed in the case of the GRS 1915+105 observations in \S \ref{ssFitting}.
\vspace{0.2in}
%------------------------------------------------------------------------------
%****************************************************************************** %------------------------------------------------------------------------------

\section{Application to GRS 1915+105}\label{sObs}

\subsection{Observations}\label{ssObs}

\begin{figure}
\centering
\includegraphics[width=8cm]{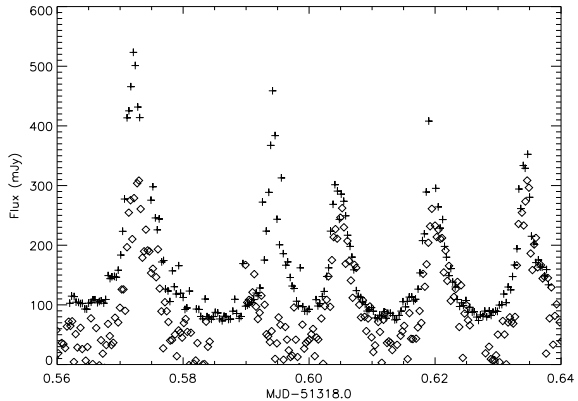}
\caption{Infrared 2.2 $\mu$m data (crosses) from the IRCAM3 instrument on UKIRT, and millimetre 1.3 mm data (diamonds) from the SCUBA instrument on JCMT. Taken from Fender \& Pooley (2000).}
\label{fData}
\end{figure}

\begin{figure}
\centering
\includegraphics[width=8cm]{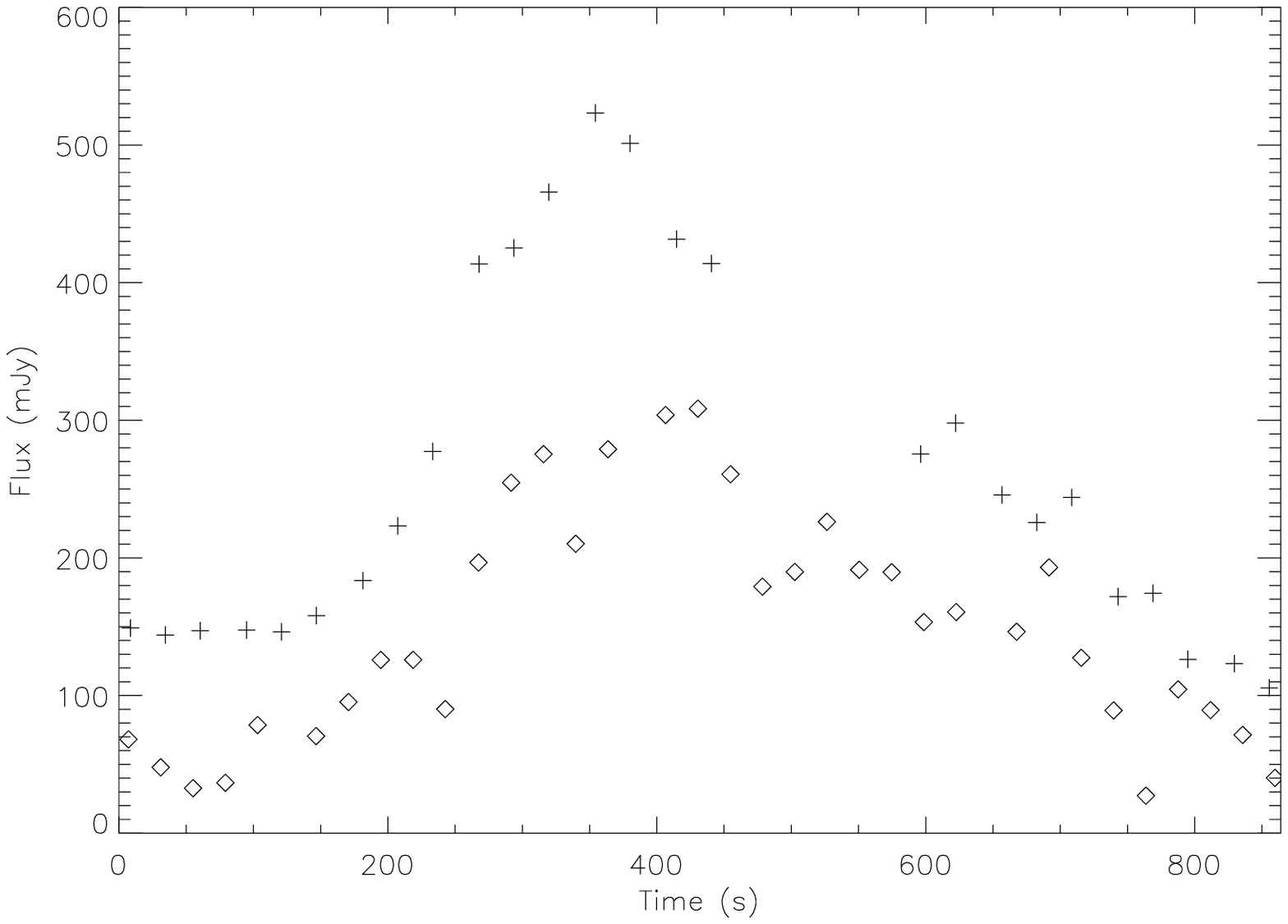}
\includegraphics[width=8cm]{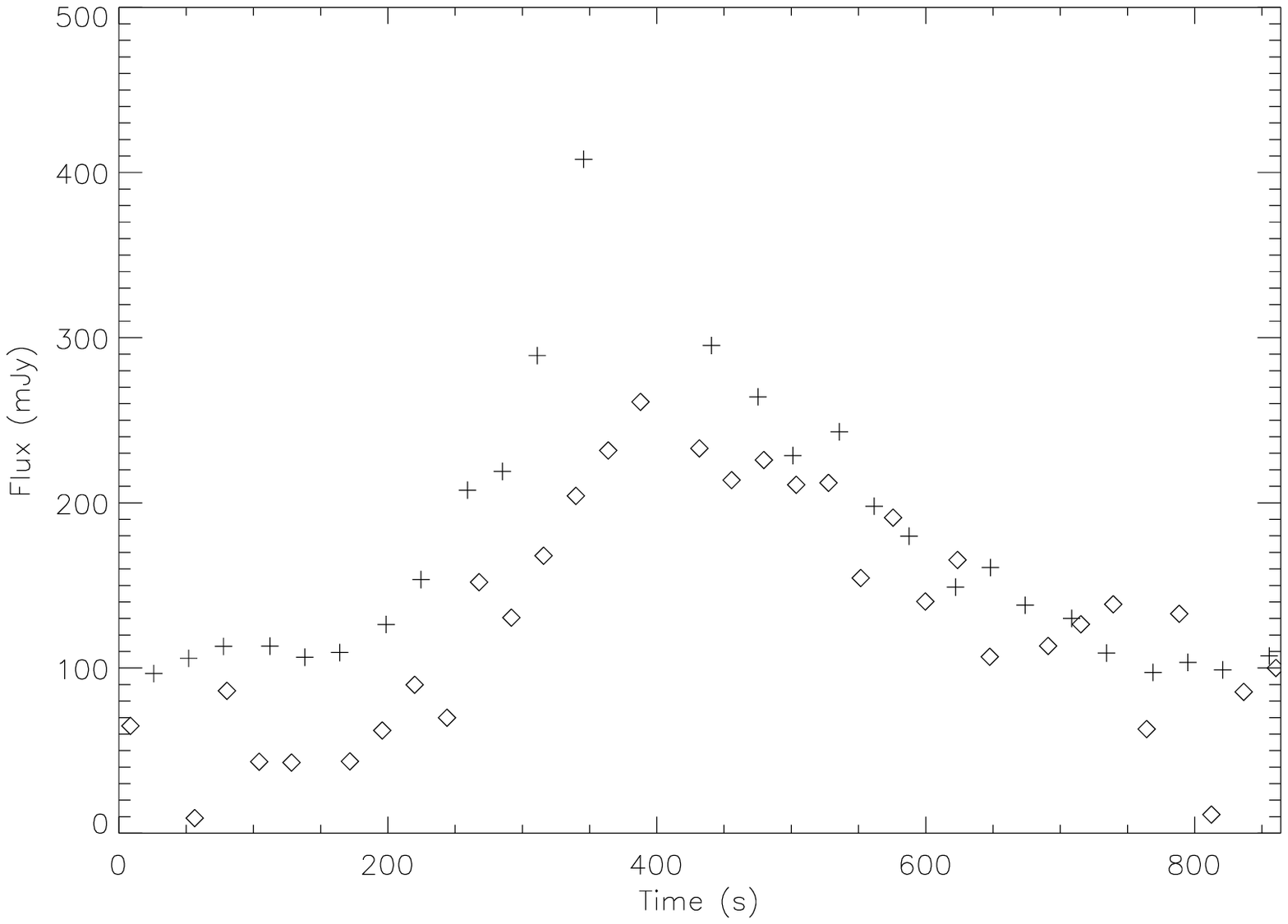}
\caption{Detailed view of a) 1st flux peak b) 4th flux peak in the data shown in figure \ref{fData}}
\label{fPeaks}
\end{figure}

GRS 1915+105 was observed simultaneously at two wavelengths on 1999 May 20 (Fender \& Pooley 2000); at 2.2 $\mu$m with the IRCAM3 instrument on the United Kingdom Infrared Telescope (UKIRT), and at 1.3 mm with the SCUBA instrument on the James Clerk Maxwell Telescope (JCMT). The resulting light curves recorded over a two hour period are shown, overlaid, in figure \ref{fData}. The light curves show large amplitude quasi-periodic oscillations, each lasting $\sim 1000$ s, believed to be ejection events in the jet. Detailed views of the first and fourth peaks are shown in figure \ref{fPeaks}. 

The infrared data from UKIRT has been dereddened with an assumed infrared $K$-band extinction of $A_K = 3.3$ mag. However, there is an uncertainty in the infrared flux values of at least 40 per cent (Fender et al. 1997) as the precise value for the absorption correction is not known. Furthermore we expect that the infrared flux will have some unknown background contribution due to the emission from regions of the GRS 1915+105 system other than the jet. The error values on individual data points are of the order of 5 per cent (Fender, private communication). 

The millimetre data show flux peaks that occur almost simultaneously with the infrared peaks. The time lag between the millimetre and infrared peaks is less than 50 s, though the sampling rate is not sufficient to provide a more exact determination. In the case of the second infrared peak there is no corresponding peak in the millimetre data. This peculiarity could prove useful in understanding the physics behind these oscillations (see \S \ref{ssFlat}).

\subsection{Properties of GRS 1915+105}\label{ssProps}

We can use observations of GRS 1915+105 to set the fixed parameters of our model described in \S \ref{sBKTEModel}. These parameters are summarised in table \ref{tGRS}. As explained in \S \ref{sBKModel}, the inclination of the jet is implicitly assumed by the model to be $i = 90^{\circ}$, which is in reasonable agreement with observed inclinations of $i \approx 70^{\circ}$ (Mirabel \& Rodr\'{\i}guez 1994). The opening half-angle of the jet, $\theta_{\rm j}$, is observed to be less than $4^{\circ}$ (Fender et al. 1999), so we take $w_0 = 0.05 \, r_0$. For the distance to GRS 1915+105, $D_{\rm j}$, we assume 11 kpc (Fender et al. 1999). Following Kaiser et al. (2000) we adopt the value of $v_{\rm j} = 0.6 c$ for the bulk velocity of the electron plasma in the jet. We neglect any relativistic effects on the emission spectra as these will be small for the adopted parameters (see \S \ref{ssFlux}).

We assume that the radius of the initial shock acceleration, $r_{\rm min}$, is where the relativistic electrons are first injected and thus is limited to the size of the last stable orbit for a non-rotating black hole, which is approximately three times the Schwarzschild radius, $r_{\rm s}$. Taking the lower mass limit for the GRS 1915+105 black hole of 10 M$_\odot$ (Greiner, Cuby \& McCaughrean 2001), gives the lowest plausible value for $r_{\rm min} = 3 r_{\rm s} = 1.0 \times 10^{5}$ m. The value of $t_{\rm max}$ is assumed to equal the rise time of the observed infrared flux peaks which is $\sim 350$ s.

\begin{table}
  \begin{center}
  \begin{tabular}{c c c c c}
    \hline
    $w_0$ & $D_{\rm j}$ & $v_{\rm j}$& $r_{\rm min}$  & $t_{\rm max}$ \\
    \hline
     0.05 $r_0$ & 11 kpc & 0.6 c & $1.0 \times 10^5$ m & 350 s\\
    \hline
  \end{tabular}
  \end{center}
\caption{Fixed parameters of the jet model, determined by observations of GRS 1915+105.}
\label{tGRS}
\end{table}

\begin{figure}
\centering
\includegraphics[width=8cm]{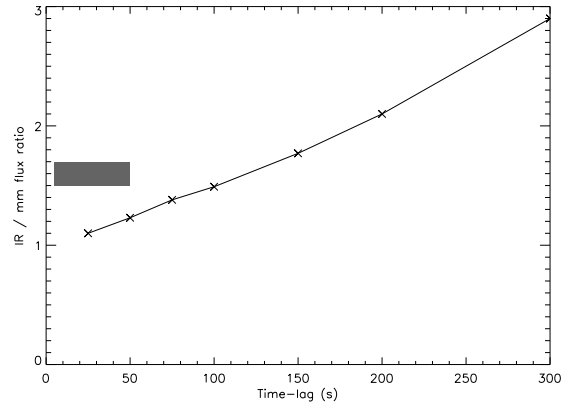}
\caption{The predicted time lag and flux ratios between the flux peaks at $1.4 \times 10^{14}$ Hz and at $2.3 \times 10^{11}$ Hz for $10^{3} \leq a_0 \leq 3.5 \times 10^{8}$, where $r_0 = 1.0 \times 10^{11}$ m. The grey rectangle shows the region where the predicted relationship would agree with the data.}
\label{f_dTvsFR}
\end{figure}

\subsection{Light curve modelling}\label{ssLCModel}

The light curves seen from the GRS 1915+105 observations seem to be best fit by the Gaussian electron density injection function. However, the light curve profile reveals a rise period that is apparently shorter than the decay period, suggesting a less symmetric functional form of $k_0^{\prime}(t)$ and hence disallowing a direct comparison between the model and data. Despite this, the model may still be used to study the time lags between the infrared and millimetre flux peaks, as this is a property of the light curve which is independent of the injection function profile (see \S \ref{sBKTEModel}).

The longer decay period may be suggestive of a delay between the time of maximum injected electron density and the time of the infrared flux peak. This may indicate the presence of a flat spectrum extending beyond the infrared into optical frequencies. However, for a fixed emission region size the extent of the flat spectrum beyond the infrared frequency of observation is solely determined by the value of $a_0$, which is constrained by the observed time lags between flux peaks. Therefore, the light curve profiles are most probably due to an electron density injection profile effect.

Using the model parameters of GRS 1915+105 shown in table \ref{tGRS}, we performed several simulations of the predicted light curves for the infrared ($1.4 \times 10^{14}$ Hz) and the millimetre ($2.3 \times 10^{11}$ Hz) fluxes with different values of the $a_0$ parameter. The resulting relationship between the time lag and the flux ratio between the flux peaks of the two frequencies is displayed in figure \ref{f_dTvsFR}. Acceptable combinations of the time lag and the flux ratio allowed by the data are denoted by the grey rectangle. Obviously, the model does not agree with the data. Modelling smaller jet opening angles simply shifts the data points for a given value of $a_0$ down to a lower position upon the same relationship line. The flux ratio for a given time lag may be increased to agree with the data if the velocity of the jet was lowered significantly, but this is in strong disagreement with our understanding of the relativistic jet. Finally, even substantial changes to the value of $r_{\rm min}$ have only a small effect on the relationship.

\begin{figure}
\centering
\includegraphics[width=8cm]{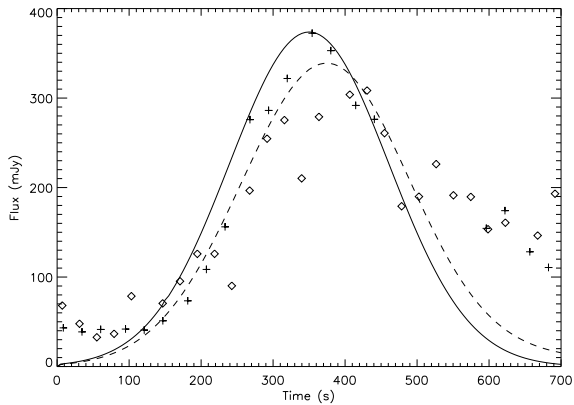}
\caption{The model of the first infrared/millimetre peak as fitted to the observed peak fluxes and time lag, with the data overlaid using the same symbols as in figure \ref{fData}. A better fit of the profile could be achieved by altering the electron density injection function, and by manipulating the balance between the background and extinction corrections.}
\label{fFit}
\end{figure}

If our model is correct, then this finding suggests that the infrared flux data may not represent the true infrared flux from the jet, due to non-jet related background emission and perhaps also an overestimate of the extinction (see also Fender et al. 1997). Therefore the true flux ratio may well be smaller. On the basis of this premise we now attempt to fit the model to the data, using the predicted flux ratio that corresponds to the observed time lag to determine the true infrared flux. We then check whether the thus determined infrared flux is within the observational uncertainty. If so, we can then determine the parameters $k_0$ and $B_0$ from the model.

\subsection{Data fitting}\label{ssFitting}

The observed time lag between the flux peaks of the two frequencies is approximately 25 s. The value of $a_0$ that produces this time lag is $1.0 \times 10^{3}$ (with $r_0 = 1.0 \times 10^{11}$ m), for which our model predicts a flux ratio between the two peaks of 1.1. For the first infrared peak in the data, the corresponding millimetre peak has a flux of $\sim 340$ mJy, so from our model we would then expect the infrared peak to have a flux of $\sim 375$ mJy. As the observed infrared peak has a flux of $\sim 525$ mJy in the data, we correct every data point by subtracting a hypothetical background emission component of 100 mJy, and then scale by a factor of 0.88 due to the uncertain extinction value. The same flux ratio may be achieved with a simple subtraction of a 150 mJy background component, however the quiescent infrared flux in-between flux peaks suggests that this value is too large. The reduced extinction is well within the 40 per cent flux uncertainty (see \S \ref{ssObs}).

Figure \ref{fFit} shows the model fit to the millimetre peak flux which obtained a value for the magnetic field strength at the radius of initial shock acceleration, $r_{\rm min} = 1.0 \times 10^{5}$ m, of $B(r\!=\!r_{\rm min}) = 7.8 \times 10^{-7}$ T, which corresponds, for the model value of $a_0$, to a peak injected electron density of $k(r\!=\!r_{\rm min}) = 5.4 \times 10^{40}$ m$^{-3}$. 

\begin{figure}
\centering
\includegraphics[width=8cm]{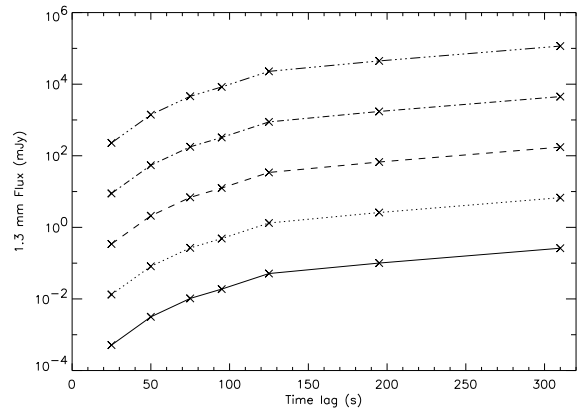}
\caption{The predicted flux from 1.3 mm emission as a function of the time lag between the infrared and millimetre flux peaks. Each line represents a different maximum injected electron density, from the bottom line of $k(r\!=\!r_{\rm min}) = 10^{16}$ (solid line), $10^{22}, 10^{28}, 10^{34}, $ to $10^{40}$ (dashed line) m$^{-3}$ at the top. The time lags are solely determined by the values of $a_0$.}
\label{f_dTvsFmm}
\end{figure}

This extreme value for the electron density cannot be avoided as is shown in figure \ref{f_dTvsFmm}. This plot shows the range of peak millimetre fluxes as a function of time lag for different values of $k(r\!=\!r_{\rm min})$, and $B(r\!=\!r_{\rm min})$. It is clearly impossible to obtain millimetre flux levels of the order of $\sim 100$ mJy without extremely high electron densities, for time lags of less than 300 s. By using this time variability analysis we have avoided the need to rely upon the uncertain flux level of the infrared emission. Furthermore, this is not just a problem with this time dependent model, but is a fundamental problem of the BK79 model, as can be seen by applying the results of \S\ref{ssBKModel_Numerical} to the observations of GRS 1915+105. 

Finally, to maintain the emission of the highest frequency radiation at the outer edge of the jet, the electron energy distribution must extend to a Lorentz factor of $\gamma_{\rm max} = 10^{7}$ for our fitted value of the magnetic field strength. Although this value is high there is some evidence, from AGN jet observations (Dermer \& Atoyan 2002), that electrons in jets can be accelerated to such energies. However, this high value of $\gamma_{\rm max}$ is a direct consequence of the low fitted value of the magnetic field strength, which would be higher in a model that does not require extremely high electron densities.

%------------------------------------------------------------------------------
%****************************************************************************** %------------------------------------------------------------------------------

\section{Discussion}\label{sDis}

\subsection{Summary}\label{ssSum}

In this paper we have demonstrated how the BK79 model produces a flat synchrotron spectrum from a conical jet geometry following the assumption that the emitting electron plasma is free of energy losses as it expands away from its origin. We have shown that the frequency range covered by this flat spectrum depends upon the radial extent of the emission region with respect to the radius where the electron plasma becomes optically thin to radiation at each frequency. This radius is determined by the value and radial dependence of the magnetic field strength and electron density within the jet. From observations of the frequency range that the flat spectrum extends over, and the observed flux from these frequencies, both the magnetic field strength and the electron density in the jet may be individually determined. 

We also adapted the BK79 model to include time variability. This not only allows comparisons with light curve observations, and places stronger demands upon the hitherto unexplained re-energisation process, but also places more stringent constraints upon the model. Prior to the inclusion of time variability the extent of the emission region had to be assumed. The time scale of variability now sets the spatial extent of the emission region. Furthermore the model parameters may be derived from observation of the time lag between peaks at each frequency rather than the less reliable measure of the flux ratio between the peaks at each frequency. Time lags are caused by an optical depth effect that occurs whether or not the flat spectrum extends beyond the observed frequency range, and hence time lags are a more reliable indicator of the physics of the system than the flux ratio. The predicted time lag would be the same even with an inverted spectrum that fully covers the observed frequency range.

This time variability model was then applied to observations of the GRS 1915+105 microquasar, which revealed a discrepancy between the model light curve profile and the observed profile. Furthermore, the model calculated for the known parameters of GRS 1915+105 required extremely large electron densities to explain the observed flux. As we believe that the system parameters are sufficiently well constrained, the only conclusion must be that the model itself is flawed. 

\subsection{On the flux discrepancy}\label{ssFlux}

As demonstrated in figure \ref{f_dTvsFmm}, we have shown that the observed millimetre flux levels are only predicted by the models that also predict time lags, between the infrared and millimetre peaks, which are many times greater than those observed (for physically justifiable electron density values). We believe that the observed millimetre flux levels are an accurate representation of the true millimetre jet emission, with no significant background contamination, as the emission drops to zero between flux peaks. Therefore, the model is fundamentally incapable of predicting millimetre flux levels of the order of 100 mJy, together with time lags between flux peaks that are less than 100 seconds, which is what has indisputably been observed.

The bulk gas flow within the jet is relativistic. Therefore, if relativistic effects can cause the observed time lag to be considerably shorter than the time lag in the rest frame of the jet material, then the observed flux will be predicted from a lower electron density. Although time dilation has the opposite effect, relativistic Doppler shifts can in principle cause this effect. However, in the case of the GRS 1915+105 jet, the inclination of the jet axis to our line-of-sight is well constrained to $\sim 70^{\circ}$, and for such inclinations the maximum effect on the observed flux from relativistic Doppler shift and Doppler beaming is only 20 per cent. To explain the observations with believable electron densities the predicted flux needs to be $10^{4}$ times greater. Such a difference cannot easily be rectified by geometrical effects, nor by the current uncertainty in the distance to GRS 1915+105. Furthermore, the simplification in using the equivalent line-of-sight optical path length for a jet at an inclination of $90^{\circ}$ does not significantly effect the results. Only for substantially smaller viewing angles and highly relativistic jet velocities do relativistic effects lead to a significant shortening of observed time lags compared to the jet rest frame.

The observations are almost certainly of synchrotron emission, as thermal emission would require even greater electron densities (Dhawan et al. 2000), and polarisation observations agree with this assertion (e.g. Fender et al. 2002). The flux discrepancy cannot be explained by the presence of the more extended jet beyond the region of the time variability, as the time variability still has to explain a large flux increase. 

In conclusion it is impossible to explain synchrotron emission of this strength from a BK79 type jet with justifiable electron densities, when restricted to the observed time lags.

\subsection{Is the flat spectrum flat?}\label{ssFlat}

From a sample of just two frequencies we cannot conclusively say that the observed flux ratio is due to a flat spectrum which does not completely extend to the lower frequency. The other possibility is that the 'flat' region of the spectrum extends to cover both of the observed frequencies and has a slightly positive slope. Such a scenario would naturally arise if the energy losses in the jet due to adiabatic decompression were partially included. However, if adiabatic losses are fully included then the resulting inverted spectrum would have a slope that is significantly steeper than that observed. The steady-state jet model of Falcke (1996) includes partial energy losses within the jet, and, for the GRS 1915+105 jet inclination angle, predicts a spectral slope of $\alpha \approx 0.2$. The observations allow a spectral slope of $0.0 \leq \alpha \leq 0.12$, due to the uncertainty on the infrared flux, so we cannot exclude the possibility of a fully extended, truly flat spectrum. 

Hence, there are two possible scenarios to explain a non-unity flux ratio between the flux peaks of the two observed frequencies. Either the lower frequency observation is of the optically thick region (or the transition to this region) of the spectrum from a BK79 type jet, or it is of the 'flat' region of a jet with energy losses. There exists a simple observable difference between these two scenarios; whereas for the former case a relationship exists between the time lag and the flux ratio, for the latter case these two parameters are independent of each other. Although the loss of the relationship between the time lag and flux ratio will not in itself affect the predicted electron densities, its implication of energy losses within the jet could lead
to a slightly lower predicted electron density. 

The missing millimetre flux peak to coincide with the second observed infrared peak does seem to suggest that for the second ejection event the flat spectrum does not extend all the way down to the millimetre waveband. However, this explanation is applicable to either of the scenarios discussed above, and is perhaps more likely than a shift in the spectral slope.

\subsection{Conclusion}\label{ssConc}

We have developed a time-dependent version of the partially self-absorbed jet model of Blandford \& K{\"o}nigl (1979). As expected from the original steady-state model, the time-dependent model gives rise to flat broadband spectra but cannot explain the large flux variations observed in GRS 1915+105 at millimetre and infrared frequencies without invoking unrealistically high electron densities. This result also holds true if the time variability is ignored. In this case the parameters of the steady-state model are determined through the observed flux ratio of the two frequencies at any given time, rather than from the observed time lag between flux peaks of the two frequencies. Therefore either this variable emission does not originate from the jet of GRS 1915+105, or an alternative model is required to explain microquasar jet emission.

Observations of the jet spectrum and time variability at different frequencies will help resolve the issue of whether the spectrum is flat, and whether we are observing optical depth effects. Observing a relationship between the flux ratio and the time lag would confirm that the flux ratio is due to an optical depth effect rather than due to a mildly inverted spectrum.

Recent observations of the infrared flaring behaviour of GRS 1915+105 (Rothstein \&  Eikenberry 2002) have illustrated the complex nature of the time variability in the emission from relativistic jets. Hence, more complex models will ultimately be required to fully explain the observations.

\section*{Acknowledgements}

We would like to thank Rob Fender for providing us with the data on GRS 1915+105 in electronic format, and the anonymous referee for providing helpful comments. RSC thanks the EPSRC for financial support.

\label{lastpage}
\end{document}